\documentclass[conference]{IEEEtran}
\usepackage{amssymb}
\usepackage{algorithm}
\usepackage{algorithmic}
\usepackage{graphs}
\usepackage[german]{babel}

\makeatletter
\def\ps@headings{%
\def\@oddhead{\mbox{}\scriptsize\rightmark \hfil \thepage}%
\def\@evenhead{\scriptsize\thepage \hfil \leftmark\mbox{}}%
\def\@oddfoot{}%
\def\@evenfoot{}}
\makeatother
\begin{document}
\title{Scheduling in Multi-hop Wireless Networks with Priorities}
\author{\authorblockN{Qiao Li}
\authorblockA{qiaoli@cmu.edu\\
Department of Electrical and Computer Engineering \\
Carnegie Mellon University \\
5000 Forbes Ave., Pittsburgh, PA 15213} \and
\authorblockN{Rohit Negi}
\authorblockA{negi@ece.cmu.edu\\
Department of Electrical and Computer Engineering \\
Carnegie Mellon University \\
5000 Forbes Ave., Pittsburgh, PA 15213}} \maketitle

\begin{abstract}
In this paper we consider prioritized maximal scheduling in
multi-hop wireless networks, where the scheduler chooses a maximal
independent set greedily according to a sequence specified by
certain priorities. We show that if the probability distributions of
the priorities are properly chosen, we can achieve the optimal
(maximum) stability region using an i.i.d random priority assignment
process, for any set of arrival processes that satisfy Law of Large
Numbers. The pre-computation of the priorities is, in general,
NP-hard, but there exists polynomial time approximation scheme
(PTAS) to achieve any fraction of the optimal stability region. We
next focus on the simple case of static priority and specify a
greedy priority assignment algorithm, which can achieve the same
fraction of the optimal stability region as the state of art result
for Longest Queue First (LQF) schedulers. We also show that this
algorithm can be easily adapted to satisfy delay constraints in the
large deviations regime, and therefore, supports Quality of Service
(QoS) for each link.
\end{abstract}


\section{Introduction}
\label{sec_intro}

Efficient scheduling algorithms in wireless networks have been the
subject of intensive research in the past few years. A fundamental
issue is that the optimal scheduling algorithm, which can achieve a
superset stability region of any scheduler, involves solving a
Maximum Weighted Independent Set (MWIS) problem \cite{tassiulas92},
which is, in general, NP-Hard \cite{sharma06}. This phenomenon is
exaggerated by the requirement that the MWIS problem has to be
solved \emph{in every time slot}, which renders it impossible to be
implemented in applications due to the high computing resource
consumptions. As an alternative, low complexity suboptimal
schedulers with guaranteed efficiency ratio are needed, among which,
maximal scheduling has been the focus of recent research. Similar to
the maximal matching in the switch scheduling literature
\cite{mckeown}, maximal scheduling is a low complexity algorithm in
wireless networks with a constant efficiency ratio which is
inversely proportional to the maximum number of independent links in
a link's neighborhood \cite{lin05}, \cite{chaporkar08}. Not only is
it sound theoretically, maximal scheduling is also well supported in
practice by efficient distributed algorithms with constant overheads
(e.g. see \cite{gupta07}).

Furthermore, it has been observed that the performance guarantee for
maximal schedulers can be improved significantly by considering
specific maximal schedulers, since the class of maximal schedulers
is quite broad. Recently, LQF scheduling has been shown to yield a
much larger stability region than the worst case maximal scheduling.
During scheduling, an LQF scheduler produces a maximal independent
set greedily following a sequence according to the queue length
order of each link, from the longest to the shortest. It is shown
that in certain networks, where the topology satisfies the so called
``local-pooling'' condition \cite{dimakis05}, LQF is optimal. For
general networks, the performance of LQF scheduling can be bounded
by its ``local-pooling factor'' \cite{joo08}, which is a function of
the network topology and interference model. In the worst case, LQF
scheduling can guarantee an efficiency ratio between $1/6$ and $1/3$
in geometric networks with the $K$-hop interference model.

Promising as it is, presently there are still some open issues for
the LQF scheduling. First of all, the stability region of LQF is not
well characterized, i.e., it is unclear whether a network is stable
or not under an arrival process whose rate is outside the stability
region of the the worst case maximal scheduler. Moreover, unlike
maximal matching for switch scheduling \cite{mckeown}, where a
centralized controller is available, distributed LQF scheduling in
wireless networks is hard to implement, and is subject to
performance degradation due to asynchronous queue length information
updates. Finally, it is hard to adapt LQF scheduling to support QoS
in wireless multi-hop networks, due to the intractability of its
analysis, which allows its performance to be understood only in
certain networks with sufficient symmetries \cite{ying06}.

Realizing the limits of LQF scheduling, we try to improve the
performance of maximal scheduling from a different perspective. Note
that in LQF scheduling, in each time slot $n$, the scheduler picks
the links following a sequence according to queue lengths. We
generalize this to \emph{prioritized} maximal scheduling, where in
each time slot the scheduler chooses a maximal schedule following a
sequence specified by a priority vector ${\bf p}(n)$, which is not
necessarily the same as the queue length orders. The priority
process ${\bf p}(n)$ is, in general, a random process, and includes
the priorities generated by the max-weight scheduling
\cite{tassiulas92} and LQF scheduling as special cases. Therefore,
it can achieve the optimal stability region, if the priorities are
properly chosen. In this paper, we show that a simple i.i.d random
process ${\bf p}(n)$ suffices to achieve the optimal stability
region. Furthermore, the distribution of ${\bf p}(n)$ can be
computed, or approximated by a PTAS, if the arrival rate value
provided as of each link is available. In applications, the arrival
rates can be obtained from either online estimation or the
parameters by upper layer services, such as digital voice or encoded
video. Since arrival rates parameters and pre-computation of the
priorities are only needed when the network topology changes, this
combined priority pre-computation and maximal scheduling approach is
suitable for slowly changing wireless multi-hop networks.

Next we consider a special class of prioritized maximal schedulers,
where ${\bf p}(n)$ is constant over time. In principle, this is
simpler than the LQF scheduler, where dynamic priorities are used,
and its performance is not influenced by asynchronous queue length
updates. Interestingly, we show that if the priorities are properly
chosen, we can achieve \emph{the same} efficiency ratio as LQF
scheduling. Moreover, we give a specific lower bound
characterization of the stability region of prioritized maximal
scheduling with a fixed ${\bf p}$. Further, we provide an algorithm
that can successfully compute a stabilizing priority ${\bf p}$ as
long as the arrival rate vector is inside the lower bound region of
\emph{any} fixed priority scheduler. Therefore, it is easy to check
the stability of the prioritized maximal scheduler for any arrival
process that is within a certain guaranteed fraction of the optimal
stability region.

Finally we consider the delay constrained scheduling problem in
wireless networks. With a major class of services being real time
services, which are sensitive to congestion in the network causing
buffer overflows, the issue of determining and guaranteeing delay in
wireless networks is of equal, if not more, importance compared to
the issue of stability. It is desirable to implement a scheduling
algorithm such that the queue overflow probability for each link is
below a certain small threshold. In this paper we try to solve this
problem by using maximal scheduling with constant priority. We first
analyze delay guarantees of the worst case maximal scheduler, as
well as those of the prioritized maximal scheduler in the large
deviations regime, and formulate an upper bound on the queue
overflow probability. By exploring the similarity between the
stability constrained and the delay constrained scheduling problems,
we then propose a greedy priority assignment algorithm, which is
adapted from the priority assignment algorithm designed for the
stability case, with the guarantee that it will generate a
satisfying priority ${\bf p}$ as long as the delay constraint can be
achieved by some priority.

The paper is organized as follows. Section \ref{sec_model}
introduces the system model. Section \ref{sec_random} describes the
prioritized maximal scheduler with random priorities. We consider
the constant priority case in Section \ref{sec_constant}, and adapt
it to support delay constraints in Section \ref{sec_delay}. Section
\ref{sec_conclusion} concludes this paper.

\section{System Model}
\label{sec_model}

\subsection{Network Model}

\begin{figure}[t]
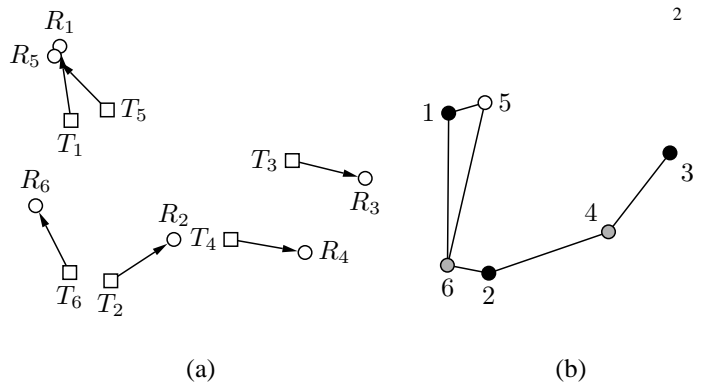

\begin{center}
\begin{tabular}{c c}
\begin{graph}(4.5,4.5)(.2,.2)
\graphnodecolour{1} \grapharrowlength{.2} \filledareasfalse
\roundnode{R1}(0.5728,4.1807) \autonodetext{R1}[n]{$R_1$}
\roundnode{R2}(2.0898,1.6090)\autonodetext{R2}[n]{$R_2$}
\roundnode{R3}(4.6345,2.4221) \autonodetext{R3}[s]{$R_3$}
\roundnode{R4}(3.8327,1.4362) \autonodetext{R4}[e]{$R_4$}
\roundnode{R5}(0.5042,4.0461) \autonodetext{R5}[w]{$R_5$}
\roundnode{R6}(0.2527,2.0587) \autonodetext{R6}[n]{$R_6$}
\squarenode{T1}(0.7220,3.1919)\autonodetext{T1}[s]{$T_1$}
\squarenode{T2}(1.2531,1.0613)\autonodetext{T2}[s]{$T_2$}
\squarenode{T3}(3.6635,2.6611)\autonodetext{T3}[w]{$T_3$}
\squarenode{T4}(2.8478,1.6092)\autonodetext{T4}[w]{$T_4$}
\squarenode{T5}(1.2041,3.3319)\autonodetext{T5}[e]{$T_5$}
\squarenode{T6}(0.7066,1.1677)\autonodetext{T6}[s]{$T_6$}
\diredge{T1}{R1} \diredge{T2}{R2} \diredge{T3}{R3} \diredge{T4}{R4}
\diredge{T5}{R5} \diredge{T6}{R6}
\end{graph}
&
\begin{graph}(4.5,4.5)(.1,.1)
\graphnodesize{.2}\graphnodecolour{1} \grapharrowlength{.25}
\filledareasfalse \roundnode{L1}(0.7220,3.1919)[\graphnodecolour{0}]
\autonodetext{L1}[w]{$1$}
\roundnode{L2}(1.2531,1.0613)[\graphnodecolour{0}]
\autonodetext{L2}[s]{$2$}
\roundnode{L3}(3.6635,2.6611)[\graphnodecolour{0}]
\autonodetext{L3}[se]{$3$}
\roundnode{L4}(2.8478,1.6092)[\graphnodecolour{.7}]
\autonodetext{L4}[nw]{$4$} \roundnode{L5}(1.2041,3.3319)
\autonodetext{L5}[e]{$5$}
\roundnode{L6}(0.7066,1.1677)[\graphnodecolour{.7}]
\autonodetext{L6}[s]{$6$} \edge{L1}{L5} \edge{L1}{L6} \edge{L2}{L4}
\edge{L2}{L6} \edge{L3}{L4} \edge{L5}{L6}
\end{graph}
\\
(a)&(b)
\end{tabular}
\end{center}
\caption{\label{fig:sample}(a) is an arbitrary network of $6$ links,
and (b) is its interference graph. The black nodes form a {\em
maximum }independent set, while the gray nodes form a {\em maximal}
independent set.}
\end{figure}

We model the topology of the network as a directed graph $G=(V, E)$,
where $V$ is the set of user nodes, and $E$ is the set of
communication links. A link $i=(u, v)\in E$ only if node $v$ is in
the transmission range of node $u$. The interference is modeled by
an undirected interference graph $G_c=(V_c, E_c)$, where $V_c$ is
the set of links, and $E_c$ is the set of pairwise conflicts. Two
links $(i, j)\in E_c$ if and only if they are not allowed to
transmit together. For example, in the \emph{primary} interference
model, the only constraint is that a user node can not transmit and
receive simultaneously. Therefore, two links $(i, j)\in E_c$ if and
only if they share a common node, in which case the scheduling
problem is reduced to a \emph{matching} problem in $G$. This model
arises naturally in switch scheduling, and is suitable for wireless
networks using Bluetooth or FH-CDMA physical layers \cite{lin05}. If
\emph{secondary} interference is considered, one model is the
$K$-hop interference model \cite{sharma06}, which requires that two
links within $K$ hops can not transmit at the same time (Note that
the 802.11 DCF (Distributed Coordination Function) corresponds to
$K=2$). See Fig. \ref{fig:sample} for an illustration of the network
topology and interference graph for a sample network consisting of
$6$ links.

For each link $i$, define its neighbor set as $N_i=\{j: (i,j)\in
E_c\}$. Thus, the transmission of link $i$ is successful if and only
if no link in $N_i$ is transmitting. Denote the interference degree
of link $i$ as $\Delta_i$, which is the cardinality of the maximum
independent set in the subgraph formed by $\{i\}\cup N_i$. It has
been shown that $\Delta=\max_{i\in V_c}\Delta_i$ is related to the
efficiency ratio of maximal scheduling \cite{chaporkar08}. To
analyze the performance ratio of the prioritized maximal scheduling,
we need the following definitions. Denote a node removal sequence as
${\bf r}=(i_1, i_2,\ldots, i_{{|V_c|}})$, which is a permutation of
$(1, 2, \ldots, |V_c|)$, and consider removing the nodes from $G_c$
according to the sequence $\bf r$. Thus, $i_k$ denotes the index of
the $k$th node that is removed. If a certain node $i$ is removed in
some step $k$ (i.e., $i_k=i$), define $\delta_i^{({\bf r})}$ as the
interference degree of link $i$ when it gets removed in the subgraph
at that step (i.e., without nodes $i_1, i_2, \ldots, i_{k-1}$).
Define\vspace{-.1in}
\begin{eqnarray*}
\delta=\min_{{\bf r}\in \Pi}\max_{i\in V_c}\delta_i^{({\bf r})},
\end{eqnarray*}
\vspace{-.2in}

\noindent where $\Pi$ is the set of $n!$ removal (permutation)
sequences. It has been shown that $1/\delta$ is a lower bound on the
local pooling factor for a network \cite{joo08}, and hence, the
lower bound on the efficiency ratio of the LQF scheduling. We will
show later that $\delta$ is also a lower bound on the efficiency
ratio of maximal scheduling with constant priority.

\subsection{Traffic Model}

We assume the network is time synchronized, and that the transmitter
of each link $i$ is associated with an external arrival process
$A_i(n)$, which is the cumulative number of packet arrivals during
the first $n$ time slots. We further assume that the arrivals happen
at the end of each time slot, and in each time slot, the packet
arrivals are bounded above by a constant. The arrival processes are
also subject to the Strong Law of Large Numbers (SLLN),
i.e.,\vspace{-.05in}
\begin{equation}
\lim_{n\rightarrow\infty} A_i(n)/n = a_i\textrm{ for all }i\in V_c,
\end{equation}\vspace{-.2in}

\noindent with probability $1$ (w.p.1), where the constant $a_i$ is
the arrival rate for link $i$. Note that these assumptions on the
arrival process are quite mild, as the arrivals can be dependent
over time slots and also among different links. As a consequence, in
this paper we only focus on single hop traffic, since it is
straightforward to generalize these results to the multi-hop case
(since departures from one queue may very well be arrivals to
another queue, in our mild assumption).

\subsection{Scheduling Model}

In each time slot, a scheduler $\pi$ chooses an independent set for
transmission. In this paper, we are interested in maximal
scheduling, where a scheduled set of links has the property that no
other link can be added without violating the interference
constraints. Denote the family of maximal independent sets as
$\mathcal{M}$. We will also treat a maximal independent set ${\bf
m}\in \mathcal{M}$ as a column vector such that $m_i=1$ if $i\in
{\bf m}$, and $m_i=0$ otherwise, as long as there is no confusion.
The queueing dynamics at each link can be expressed
as\vspace{-.05in}
\begin{equation}\label{eqn: queueing}
Q_i(n) = Q_i(0)+A_i(n)-D_i(n)
\end{equation}\vspace{-.2in}

\noindent where $Q_i(n)$ is the queue length at link $i$ at the end
of time slot $n$ and $D_i(n)$ is the cumulative departures during
the first $n$ time slots. Clearly, $D_i(n)$ must ensure that
$Q_i(n)\geq 0$.

For prioritized maximal schedulers, during scheduling in time slot
$n$, an independent set is produced, as described above, guided by a
priority vector ${\bf p}(n)$, where $p_i(n)$ represents the priority
of link $i$. We assume that link $i$ has higher priority than link
$j$ if and only if $p_i(n)<p_j(n)$ and that all priorities are
distinct. Thus, the scheduler chooses departures following a
sequence specified by $\bf p$, from the highest to the lowest, and
schedules link $i$ for departure if it has a nonempty queue and no
higher priority neighbor has already been scheduled.

The throughput of a scheduler $\pi$ is represented by its stability
region $\mathcal{A}_\pi$, which is the set of stable arrival rate
vectors under $\pi$. We define stability to be rate stability,
i.e.,\vspace{-.05in}
\begin{equation}
\lim_{n\rightarrow\infty} A_i(n)/n = \lim_{n\rightarrow\infty}
D_i(n)/n = a_i\quad\textrm{w.p.1}
\end{equation}\vspace{-.2in}

\noindent for all $i\in V_c$. It has been shown that the max-weight
scheduler in \cite{tassiulas92} can achieve the maximum stability
region, $\mathcal{A}_{\max}$, which is the interior of the convex
hull of the maximal independent sets in $G_c$ (we do not consider
the boundary points in this paper). We begin our study of
prioritized maximal schedulers by considering the stability region
of prioritized maximal schedulers with random priorities, in the
next section.

\section{Scheduling with Random Priorities}
\label{sec_random}

In this section, we analyze the stability region of prioritized
maximal schedulers when the priority vectors $\big({\bf p}(n)\big)$
form a random process. First of all, as a subclass of maximal
schedulers, any prioritized maximal scheduler can support the
following region \cite{chaporkar08}:\vspace{-.05in}
\begin{equation}
\mathcal{A}_{\min}=\{{\bf a}:a_i+\sum_{j\in N_i}a_j<1, 1\leq i\leq
|V_c|\}.
\end{equation}\vspace{-.15in}

\noindent In order to obtain a better bound, we need to explore the
impact of the priority vectors. In fact, the stability region of a
priority scheduler is closely related to the choice of priorities.
For example, consider a star shaped interference graph, with center
link $1$ and $n-1$ outer links. If we assign the priorities such
that the center link $1$ always has the lowest priority, then for
any arrival rate $a_1+\sum_{j=2}^{n} a_j>1$, we can find an arrival
process which makes link $1$ unstable. Therefore, the stability
region coincides with $\mathcal{A}_{\min}$. As another example,
consider the following choice of priorities: in each time slot, we
first solve the MWIS problem\vspace{-.07in}
\begin{equation}\label{eqn: np-hard}
{\bf m}^\star(n)=\textrm{argmax}_{{\bf m}\in \mathcal{M}}{\bf
Q}(n)^T\bf{m}.
\end{equation}\vspace{-.22in}

\noindent Then, we assign all links with $m_i^\star(n)=1$ the
highest priorities (the order does not matter) while the links with
$m_i^\star(n)=0$ are assigned the remaining (lowest) priorities (the
order does not matter). Since ${\bf m}^\star(n)$ is an independent
set, all links in the independent set have the local highest
priority, i.e., the highest priority in that specific neighborhood.
After scheduling, the scheduled links maximize the weighted sum
(\ref{eqn: np-hard}), and thus $\mathcal{A}_{\max}$, in this
example, can be achieved \cite{tassiulas92}. However, the
computation of priorities is prohibitive, since (\ref{eqn: np-hard})
requires solving an NP-hard problem \emph{in every time slot}. In
fact, a much simpler prioritized scheduler can also achieve
$\mathcal{A}_{\max}$, by utilizing only an i.i.d random process
$\big({\bf p}(n)\big)$.

We need the following lemma.

\emph{Lemma 1:} For any arrival rate vector $\bf a$ and an i.i.d
random process $\big({\bf p}(n)\big)$ of priority vectors, if there
exists a $S=(S_1, S_2,\ldots, S_n)$ where set $S_i\subseteq N_i$,
such that
\begin{equation}
a_i+\sum_{j\in N_i/S_i}a_j<\Pr(p_i<p_j,\forall j\in S_i),
\end{equation}
$\bf a$ is stable under the maximal scheduler with process
$\big({\bf p}(n)\big)$.

\emph{Proof:} See in Appendix.\hfill$\blacksquare$

Thus, this lemma provide a lower bound on the stability region
achievable by $\big({\bf p}(n)\big)$. The following is the main
theorem of this section. It uses Lemma 1 to show that any ${\bf
a}\in\mathcal{A}_{\max}$ can be stabilized by a properly chosen
$\big({\bf p}(n)\big)$.

\emph{Theorem 1:} In an arbitrary network with $n$ nodes, for any
${\bf a}\in \mathcal{A}_{\max}$, there exists a stabilizing
prioritized maximal scheduler with an i.i.d process $\big({\bf
p}(n)\big)$. Furthermore, the support of $\big({\bf p}(n)\big)$
consists of at most $n+1$ elements.

\emph{Proof:} Since $\bf a$ is in an open set $\mathcal{A}_{\max}$,
there exists an $\epsilon>0$ such that ${\bf a}+\epsilon {\bf
e}\in\mathcal{A}_{\max}$, where ${\bf e}=(1, 1, \ldots, 1)^T$.
According to the Carath\'eodory theorem, ${\bf a}+\epsilon {\bf e}$
can be represented as a convex combination of at most $n+1$ maximal
independent sets, i.e.,\vspace{-.15in}
\begin{equation}\label{eqn: cvx-hull}
{\bf a}+\epsilon {\bf e}=\sum_{k=1}^{n+1} \theta_k{\bf
m}^{(k)},\quad {\bf m}^{(k)}\in\mathcal{M}
\end{equation}\vspace{-.12in}

\noindent where $\bf \theta\succeq 0$ and ${\bf e}^T\theta=1$. We
associate a priority vector ${\bf p}^{(k)}$ with each ${\bf
m}^{(k)}$ such that if $m_i^{(k)}=1$, $i$ has the local highest
priority, and other priorities are arbitrary. During the scheduling
in time slot $n$, the scheduler chooses ${\bf p}^{(k)}$ with
probability $\theta_k$. Therefore, we have\vspace{-.07in}
\begin{eqnarray*}
\Pr(p_i<p_j,\forall j\in N_i)&\geq&\sum_{k:
m^{(k)}_i=1}\theta_k\\
&=&\sum_{k}\theta_k m_i^{(k)}\\
&=&a_i+\epsilon
\end{eqnarray*}\vspace{-.25in}

\noindent and the stability follows from Lemma $1$ by setting
$S_i=N_i$ for all $1\leq i\leq n$.\hfill$\blacksquare$

Intuitively, Theorem 1 shows that the traditional coloring approach,
which is used for scheduling in networks with constant traffics, is
still applicable in the presence of stochastic packet dynamics.
Furthermore, the hardness remains almost the same, since (\ref{eqn:
cvx-hull}) is equivalent to a coloring problem. Note (\ref{eqn:
cvx-hull}) is only executed during the pre-computation phase. Thus,
compared to the max-weight scheduler, which essentially solves an
NP-hard problem in every time slot, the burden on the prioritized
maximal scheduler during the scheduling phase is significantly
relieved. Finally, we show that there exists PTAS for the
pre-computation of the priorities.

\emph{Theorem 2:} Given a network and interference model, if there
exists a PTAS for the MWIS problem, then there exists a PTAS for
computing the the stabilizing priorities.

\emph{Proof:} Suppose we have ${\bf a}+\epsilon' {\bf
e}\in(1-\epsilon)\mathcal{A}_{\max}$ for some $\epsilon'>0$. The
priorities can be obtained by solving the following
optimization:\vspace{-.1in}
\begin{eqnarray}
\min_{\bf x}&&{\bf e}^T{\bf x}\nonumber\\
\textrm{subject to}&& M{\bf x}\succeq {\bf a}+\epsilon' {\bf e},
{\bf x}\succeq 0\label{eqn: primal}
\end{eqnarray}
where $M$ is the matrix whose columns correspond to all ${\bf
m}\in\mathcal{M}$. This problem can be solved by binary search over
$t$, where in each step we assume ${\bf e}^T{\bf x}=t$ and solve the
following\vspace{-.07in}
\begin{eqnarray}
\max_{{\bf x}, s}&&s\nonumber\\
\textrm{subject to}&& {(M{\bf x})_i\over a_i+\epsilon'}\geq s\qquad 1\leq i\leq |V_c|\nonumber\\
&&{\bf e}^T{\bf x}=t, {\bf x}\succeq 0.\label{eqn: bs-primal}
\end{eqnarray}\vspace{-.2in}

By strong duality, this can be solved by the dual
problem\vspace{-.07in}
\begin{eqnarray}
\min_{\bf \lambda}&&f(\lambda)\nonumber\\
\textrm{subject to} && \lambda\succeq 0, {\bf
e}^T\lambda=1\label{eqn: dual}
\end{eqnarray}
where $f(\lambda)$ is the optimal value of the following
problem\vspace{-.07in}
\begin{eqnarray}
\max_{\bf x}&&\sum_{i=1}^n \lambda_i{(M{\bf x})_i\over a_i+\epsilon'}\qquad 1\leq i\leq |V_c|\nonumber\\
\textrm{subject to} &&{\bf e}^T{\bf x}=t, {\bf x}\succeq
0.\label{eqn: MWIS}
\end{eqnarray}\vspace{-.15in}

It has been shown that an $\epsilon$-approximate solution for
(\ref{eqn: bs-primal}) can be obtained by solving
$\Theta(\epsilon)$-approximate solutions for (\ref{eqn: MWIS})
$O(|V_c|(\epsilon^{-2}+\log|V_c|))$ times, see \cite{grigoriadis01}
for an approximation algorithm using logarithmic potential
reduction. The only thing remains is to check that (\ref{eqn: MWIS})
is a MWIS problem. Note that (\ref{eqn: MWIS}) is a linear program,
with a simplex constraint, and therefore, the optimal value is
attained at a vertex, i.e.,\vspace{-.05in}
\begin{equation}
f(\lambda)=\max_{{\bf m}\in M}{\bf w}(\lambda)^T{\bf m}
\end{equation}\vspace{-.15in}

\noindent where $w_i(\lambda)={t\lambda_i\over a_i+\epsilon}$,
showing that $f(\lambda)$ finds a MWIS. Thus, the theorem holds.
\hfill$\blacksquare$

For most network and interference models, there exists a PTAS for
the MWIS problem, see \cite{kuhn05} for an example using the graph
partitioning technique in geometric graphs. Note that one can also
achieve similar performance by using a PTAS for the max-weight
scheduler during the scheduling \cite{ray07}. Compared to their
approach, our method has the advantage that the scheduling phase has
low complexity, which is \emph{independent of approximation ratio},
since the approximation algorithm is only executed during the
priorities pre-computation phase.

\section{Scheduling with Constant Priority}
\label{sec_constant}

In this section we consider a special class of prioritized
schedulers, where ${\bf p}(n)={\bf p}$ is a constant vector, i.e.,
the priority for each link is fixed. This is certainly easier to
implement, since it does not require a global coordination variable,
specifying the priority vector applying in each time slot. We will
focus on both of the priority pre-computation and the performance
guarantees.

\subsection{Priority Assignment}

In order to search for the optimal priority, we need to analyze the
associated stability region for an arbitrary priority vector $\bf
p$. By setting $S_i=\{j\in N_i: p_i<p_j\}$ for each link $i\in V_c$
and apply Lemma $1$, we get a lower bound\vspace{-.05in}
\begin{equation}\label{eqn: fix-priority}
a_i+\sum_{j\in N_i}a_j{\bf 1}(p_i-p_j)<1\quad\forall i\in V_c
\end{equation}\vspace{-.15in}

\noindent where ${\bf 1}(\cdot)$ is the indicator
function\vspace{-.05in}
\begin{equation}\label{eqn: indicator}
{\bf 1}(x)=\left\{\begin{array}{ll} 1 & \textrm{if }x>0\\
0 & \textrm{otherwise}\end{array}\right.
\end{equation}\vspace{-.15in}

\noindent Denote this stability region as $\mathcal{A}_p$ and the
stability region achieved by \emph{any} constant priority as
$\mathcal{A}=\cup_p \mathcal{A}_p$. We are interested in computing a
stabilizing priority for a given $a\in\mathcal{A}$. We will first
propose the algorithm, and then show its optimality. See also in
\cite{li08}. Note that the proof below is new, compared to
\cite{li08}, and so is the condition that
$\mathcal{A}\subseteq\mathcal{A}'$.


\begin{algorithm}[H]
\caption{Stable-Priority $(G_c,{\bf a})$}
\begin{algorithmic}[1]
\FOR{$k=1$ to $|V_c|$}

\STATE $s\leftarrow\arg\min_{i\in V_c} \{a_i+\sum_{j\in N_i}a_j\}$;

\STATE $p_s\leftarrow n+1-k$;

\STATE Remove $s$ from $V_c$ and its incident edges from $E_c$;

\ENDFOR

\RETURN $\bf p$
\end{algorithmic}
\end{algorithm}

In each step, the algorithm chooses node $s$, the center node of a
neighborhood, and assign it the next lowest priority.

The following lemma is the key to the proof.

\emph{Lemma 2:} (\emph{Alternate definition of $\mathcal{A}$}) For
the $\mathcal{A}$ defined above, we have $\mathcal{A}=\mathcal{A}'$
where $\mathcal{A}'$ is any set of rate vectors in $\mathbb{R}^n_+$
that satisfies the following conditions:

\emph{1) (Coordinate-convex)} If ${\bf a}\in\mathcal{A}'$ and ${\bf
a}'\preceq {\bf a}$, then ${\bf a}'\in\mathcal{A}'$.

\emph{2) (Worst-case stable)} For any nonzero ${\bf
a}\in\mathcal{A}'$, there exists $1\leq i\leq n$ such that $a_i>0$
and $a_i+\sum_{j\in N_i}a_j<1$.

\emph{Proof:} Suppose ${\bf a}\in\mathcal{A}$, then there exists
${\bf p}$ such that ${\bf a}\in\mathcal{A}_{\bf p}$. Thus, property
(1) holds from the fact that $\mathcal{A}_{\bf p}$ is a polytope,
and thus contains any ${\bf a}'\preceq {\bf a}$. For this priority
${\bf p}$, denote $i$ as the lowest priority link with positive
arrival rate. Then, $S_i=\phi$, so that (\ref{eqn: fix-priority}) is
the same as property (2). Therefore we have $\mathcal{A}\subseteq
\mathcal{A}'$.

Suppose ${\bf a}^{(1)}\in\mathcal{A}'$. From property $(2)$ we
have\vspace{-.05in}
\begin{eqnarray*}
a^{(1)}_{i_1}+\sum_{j\in N_{i_1}}a^{(1)}_j<1
\end{eqnarray*}\vspace{-.12in}

\noindent for some $i_1$. Assign $p_{i_1}=|V_c|$, and form a new
rate vector ${\bf a}^{(2)}$ from ${\bf a}^{(1)}$ by setting
$a^{(1)}_{i_1}=0$ (which induces a reduced graph). From property
(1), $a^{(2)}$ is also in $\mathcal{A}$, and from property (2) there
exists $i_2\neq i_1$ such that $a^{(2)}_{i_2}>0$ and\vspace{-.05in}
\begin{eqnarray*}
a^{(2)}_{i_2}+\sum_{j\in N_{i_2}}a^{(2)}_j<1.
\end{eqnarray*}\vspace{-.12in}

\noindent We then set $p_{i_2}=|V_c|-1$ and repeat the similar
procedure until the highest priority, i.e., priority $(1)$ is
assigned. Thus, a $\bf p$ vector is obtained. We claim that $a\in
\mathcal{A}_{\bf p}$, due to the fact that
\begin{eqnarray*}
a_{i_k}+\sum_{j\in N_{i_k}}a_j{\bf
1}(p_i-p_j)=a^{(k)}_{i_k}+\sum_{j\in N_{i_k}}a_j^{(k)}<1
\end{eqnarray*}
for all $1\leq k\leq |V_c|$. Therefore we have
$\mathcal{A}'\subseteq \mathcal{A}$, and the lemma
holds.\hfill$\blacksquare$

From the above proof, it is straightforward to prove the following
theorem:

\emph{Theorem 3:} If ${\bf a}\in \mathcal{A}$, Stable-Priority will
return a $\bf p$ such that ${\bf a}\in\mathcal{A}_{\bf p}$.

\subsection{Performance Guarantee}

We next analyze the performance of the maximal scheduling with the
constant priority $\bf p$ generated by Stable-Priority. This can be
measured by the efficiency ratio $\gamma$, which is defined as
\begin{eqnarray*}
\gamma=\sup\{\sigma:\sigma\mathcal{A}_{\max}\in \mathcal{A}\}.
\end{eqnarray*}

To gain insight, we first consider the efficiency ratio of greedy
graph vertex coloring, where a sequence of vertices are removed
first and the colors are assigned in the reverse order. In fact, if
the arrival processes have equal rate, and can be approximated by
constant fluids, then the scheduling problem is reduced to a graph
coloring problem, where the rate is inversely proportional to the
number of colors used. Denote the removal sequence in a greedy
coloring algorithm as ${\bf r}=(i_1, i_2, \ldots, i_n)$. When the
nodes are colored in the reverse order, we need at most $d^{({\bf
r})}+1\doteq\max_k d_k^{({\bf r})}+1$ colors, where $d_k^{({\bf
r})}$ is the degree of node $i_k$ when it gets removed. Suppose that
the above maximum is achieved at node $i_{k^\star}$. When
$i_{k^\star}$ gets removed, its neighborhood needs at least
$\lceil{d^{({\bf r})}+1\over\delta^{({\bf r})}}\rceil$ colors (in
any optimal coloring), and at most $d^{({\bf r})}+1$ colors in the
sequential coloring. Therefore the efficiency ratio for this choice
of sequence is lower bounded by $1/\delta^{({\bf r})}$. In the worst
case, we have the efficiency ratio of $1/\Delta=1/\max_{{\bf
r}\in\Pi}\delta^{({\bf r})}$, which corresponds to the efficiency
ratio of the worst case maximal scheduler \cite{chaporkar08}. On the
other hand, if we choose the sequence properly, we can achieve
$1/\delta=1/\min_{{\bf r}\in\Pi}\delta^{({\bf r})}$. In the
following we show that the efficiency ratio $1/\delta$ can, indeed,
be achieved by $\mathcal{A}$.

\emph{Theorem 4:}
${1\over\delta}\mathcal{A}_{\max}\subseteq\mathcal{A}$.

\emph{Proof:} Denote the removal sequence which achieves $\delta$ as
${\bf r}=(i_1, i_2, \ldots, i_{|V_c|})$. For any ${\bf
a}\in\mathcal{A}$, define a sequence of arrival rate vectors as
follows:
\begin{eqnarray*}
{\bf a}^{(1)}&=&{\bf a}\\
{\bf a}^{(k)}&=&{\bf a}^{(k-1)}-a^{(k-1)}_{i_k}{\bf e}_{i_k}\quad
2\leq k\leq |V_c|
\end{eqnarray*}
where ${\bf e}_{i_k}$ is the vector that is all-zero except for an
$1$ in the $i_k$th entry. In other words, ${\bf a}^{(k)}$ is
obtained from ${\bf a}^{(k-1)}$ by setting the $i_{k}$th entry to
zero. We have for any $k$
\begin{equation}
a^{(k)}_{i_k}+\sum_{j\in N_{i_k}} a^{(k)}_j<\delta
\end{equation}
due to the fact that at most $\delta$ links's in $i_k$'s
neighborhood when it gets removed can transmit in each time slot.
Now for the priority assignment such that $p_{i_k}=n+1-k$ for $1\leq
k\leq n$, we have
\begin{equation}
a_{i_k}+\sum_{j\in N_{i_k}} a_j{\bf
1}(p_{i_k}-p_j)=a^{(k)}_{i_k}+\sum_{j\in N_{i_k}} a^{(k)}_j<\delta
\end{equation}
for all $i_k$. Thus we have ${1\over\delta}{\bf
a}\in\mathcal{A}_{\bf p}$ and the claim holds.\hfill$\blacksquare$

Interestingly, the efficiency ratio of $1/\delta$ has been shown to
also be a lower bound on the local pooling factor \cite{joo08} for
an interference graph, which is the efficiency ratio promised by LQF
scheduling. This is no coincidence, since in the special case of
constant fluid arrivals with equal rate, both belong to the family
of greedy coloring algorithms, and hence have similar performance
guarantees.

We next apply this result to certain networks and interference
models to obtain examples of worst case guarantees.

\emph{1) $K$-hop Interference:}

We first consider the $K$-hop interference model, which can be used
for a large class of networks. For instance, the ubiquitous IEEE
802.11 DCF is usually modeled as a $K$-hop interference model with
$K=2$, due to the RTS-CTS message exchanges. In the $K$-hop
interference model, two links $(i,j)\in E_c$ if and only if the
distance between one node in link $i$ (transmitter or receiver) and
one node in link $j$ (transmitter of receiver) is less than a
threshold $Kr$, where $r$ is the transmission range of a node.
Summarizing the result in \cite{joo08}, we have the following
corollary.

\emph{Corollary 1:} In a geometric graph with $K$-hop interference
model, the prioritized maximal scheduling with constant priority can
achieve an efficiency ratio between $1/6$ and $1/3$.

\emph{2) PHY-Graph:}

We next consider the PHY-Graph \cite{negi03}, which is a more
realistic interference model explicitly incorporating the physical
layer parameters, i.e., the signal-to-interference-plus-noise ratio
(SINR). In the PHY-Graph model, two links $(i,j)\in E_c$ if and only
if either of the following is true: 1) the distance between the
transmitter of link $i$ and the receiver of link $j$ is less than
$cl_j$, or 2) the distance between the transmitter of link $j$ and
the receiver of link $i$ is less than $cl_i$, where
$c=(SNR_t)^{1\over\kappa}$ is a function of the SINR threshold
$SNR_t$ and path loss exponent $\kappa$, and $l_i, l_j$ are the link
lengths of $i,j$, respectively. From their coloring bound, which is
essentially an upper bound on $\delta$, we have the following
corollary.

\emph{Corollary 2:} In a geometric graph with PHY-Graph interference
model, for fixed $\kappa$, the efficiency ratio of prioritized
maximal scheduling with constant priority is a nondecreasing
function of $SNR_t$. Particularly, when the $SNR_t$ is sufficiently
high, the efficiency ratio is bounded above by $1/7$.

\section{Delay-Aware Scheduling}
\label{sec_delay}

In this section we try to adapt the prioritized maximal scheduling
to support QoS constraints, in the form of an upper bound on
queue-overflow probability. Specifically, we assume that the system
constraint on the queue over flow probability for link $i$ is
\begin{eqnarray*}
\Pr(Q_i(0)>B_i)\leq\epsilon
\end{eqnarray*}
where $Q_i(0)$ is the stationary queue length and $B_i$ is buffer
capacity. For large $B_i$, we can assume that the buffer capacity if
infinite and approximate the queue overflow problem as the
following: calculate $\theta_i^*$, where
\begin{equation}\label{eqn: theta_i}
\theta_i^*=\liminf_{B_i\rightarrow\infty}{-1\over
B_i}\log\Pr(Q_i(0)>B_i)\geq\epsilon'.
\end{equation}
Here $\epsilon=\exp(-B_i\epsilon')$ while $\theta^*_i$ represents
the delay exponent of $Q_i$ in the large deviations regime. In order
to measure the delay performance of a scheduler $\pi$ in the large
deviations regime, using a similar notation to the stability region
$\mathcal{A}_\pi$, we define the delay region
\begin{eqnarray*}
\Theta_\pi=\{\theta\in\mathbb{R}_+^n:\theta\preceq\theta^*\}
\end{eqnarray*}
where $\theta^*$ is defined as Eqn. (\ref{eqn: theta_i}), i.e., the
set of guaranteed delay exponents under $\pi$. In contrast to the
stability region, where any the arrival process satisfying SLLN
applies, the delay region is quite sensitive to the arrival process
model, especially the burstiness of the process. Therefore, we have
to first re-define the arrival process model before analyzing the
delay region of maximal schedulers.

\subsection{Assumptions}

We need to slightly modify the arrival process model, so as to apply
the Large Deviations Principle (LDP). We now assume that the arrival
processes are independent among the links. For each link $i$, we
will follow the model in \cite{bertsimas99}. Specifically, we have
the following assumptions:

{\bf A1:} \emph{(The G"atner-Ellis theorem applies)}

For each link $i$, the log moment generation function
\begin{equation}
\Lambda_i(\theta)=\lim_{n\rightarrow\infty}{1\over
n}\log\mathbb{E}(e^{\theta A_i(n)})
\end{equation}
exists for all $\theta$, from which we can get the rate function via
the Legendre transform
\begin{equation}
\Lambda_i^*(\mu)=\sup_{\theta}(\theta\mu-\Lambda_i(\theta)).
\end{equation}

{\bf A2:} \emph{(The sample path LDP applies)}

For all $s\in \mathbb{N}, \epsilon_1, \epsilon_2>0$ and for every
scalars $b_0, b_1,\ldots, b_{s-1}$, there exists $N>0$, such that
for all $n>N$ and all $1=k_0\leq k_1\leq\ldots\leq k_s=n$,
\begin{equation}
P_i\geq e^{-n\epsilon_2+\sum_{j=0}^{s-1}(k_{j+1}-k_j)\Lambda^*(b_j)}
\end{equation}
where $P_i$ is the probability of the following event:
\begin{eqnarray*}
\{|A_i(k_{j+1})-A_i(k_j)-(k_{j+1}-k_j)b_j|\leq\epsilon_1n,0\leq
j\leq s\}
\end{eqnarray*}
Intuitively, this is the event that the arrival process is
constrained to lie within a tube around $s$ linear segments of
slopes $b_0, b_1, \ldots, b_{s-1}$, respectively.

{\bf A3:} \emph{(Convex dual analog of the sample path LDP)}

For all $s\in\mathbb{N}$, there exists $N>0$ and a function
$g(\cdot)$ with $0\leq g(x)<\infty$ when $x>0$, such that for all
$n\geq N$ and all $1=k_0\leq k_1\leq\ldots\leq k_s=n$,
\begin{eqnarray*}
\mathbb{E}(\exp(\theta^TZ))\leq\exp\{\sum_{j=1}^s(k_j-k_{j-1})\Lambda(\theta_j)+g(\theta_j)\}
\end{eqnarray*}
where $\theta=(\theta_1,\theta_2,\ldots,\theta_s)$ and $Z=(A_i(k_0),
A_i(k_2)-A_i(k_1),\ldots, A_i(k_s)-A_i(k_{s-1}))$.

It should be noted that the processes satisfying the above three
properties form a broad class, which includes most common models for
bursty traffic in realistic networks, such as renewal and
Markov-modulated processes. For detailed discussions, see
\cite{chang95}.

\subsection{Worst Case Maximal Scheduling}

We first consider the delay region of an arbitrary maximal
scheduler. In the following, we analyze the delay exponent for a
fixed link $i$. Denote the original network with a chosen arbitrary
maximal scheduler as system $\mathcal{S}$. Since it is analytically
intractable to give an exact characterization of the delay region,
we want to create a system $\mathcal{S}'$ such that 1) it is
relatively easier to analyze the delay region in $\mathcal{S}'$ and
that 2) the delay region of $i$ in $\mathcal{S}'$ is a lower bound
of that in $\mathcal{S}$. Note that due to maximal scheduling, when
link $i$ has a nonempty queue, we can guarantee that, in each such
time slot, there is at least one departure in $i$'s neighborhood
(which includes $i$). Therefore an obvious dominant system for
$\mathcal{S}$ is a server with service capacity $1$, which is fed by
one queue of queue length equal to the sum of the queues in
$\{i\}\cup N_i$. We can lower bound the delay exponent of link $i$
in $\mathcal{S}$ using the following bound based on $\mathcal{S}'$:
\begin{eqnarray*}
\Pr(Q_i(0)\geq x)\leq\Pr(Q_i'(0)+\sum_{j\in N_i}Q_j'(0)\geq x),
\end{eqnarray*}
where the RHS corresponds to the single queue in $\mathcal{S}'$
(equal to the sum of queues in $\mathcal{S}$) and can be calculated
using standard large deviations techniques. However, empirical
results show that this bound is quite loose, especially in the cases
where some neighbors in $N_i$ are likely to grow large, whereas
$Q_i$ is not. To get a better lower bound, we need to analyze the
queue length dynamics of link $i$ alone. Since we are analyzing the
worst case maximal scheduler, we assume a prioritized maximal
scheduler where link $i$ has the global lowest priority and the
other links in $N_i$ have the global highest priority, so that $i$
can not transmit if any link in $N_i$ has nonempty queue. Based on
this, we create a system $\mathcal{S}'$ consisting of a clique
formed by the nodes in $\{i\}\cup N_i$, and $i$ is assigned the
lowest priority. The priorities of the other links are arbitrary. We
will show that $\mathcal{S}'$ is a dominant system over
$\mathcal{S}$ in the following lemma. Intuitively, we are assuming
the worst case correlations among the neighbors in $N_i$, such that
the available time slot for link $i$ is minimized. This, together
with the scheduler which assigns link $i$ the lowest priority, can
be shown to dominate the original system $\mathcal{S}$ with
\emph{any} scheduler. We will prove this domination in the following
lemma.

\emph{Lemma 3:} For any $x\geq 0$ and $n\geq 0$, if
$Q_i(0)=Q'_i(0)$, we have
\begin{equation}
\Pr(Q_i(n)\geq x)\leq \Pr(Q'_i(n)\geq x)
\end{equation}
where $Q_i(n),$ and $Q_i'(n)$ are the queue lengths of link $i$ at
time slot $n$ in $\mathcal{S}$ and $\mathcal{S}'$, respectively.

\emph{Proof:} We will show that in an arbitrary sample path $\omega$
with the same arrival processes and initial queue lengths for both
systems, we have $Q_i(n, \omega)\leq Q'_i(n, \omega)$ for all $n\geq
0$. We assume both systems use the first-in-first-out policy, and
denote the arrival time and departure time of the $k$th packet at
link $i$ in $S$ as $t_k$ and $s_k$, respectively. Similarly for
$\mathcal{S}'$ we use $t_k'$ and $s_k'$. From the queueing equations
in both systems, we have
\begin{eqnarray*}
Q_i(n)&=&Q_i(0)+\sum_{k=1}^\infty {\bf
1}_{[0,n]}(t_k)-\sum_{k=1}^\infty {\bf 1}_{[0,n]}(s_k)\\
Q_i'(n)&=&Q'_i(0)+\sum_{k=1}^\infty {\bf
1}_{[0,n]}(t'_k)-\sum_{k=1}^\infty {\bf 1}_{[0,n]}(s'_k),
\end{eqnarray*}
where ${\bf 1}_{S}(x)$ is the indicator function that $x\in S$. Note
that for any sample path $\omega$, $Q_i(0,\omega)=Q'_i(0,\omega)$
and $t_k(\omega)=t'_k(\omega)$ due to the assumptions. Thus to show
that $Q_i(n)\leq Q'_i(n)$ w.p.1 for all $n\geq 0$, it is sufficient
to show that $s_k(\omega)\leq s'_k(\omega)$ for all $k\geq 1$ and
$\omega$. For simplicity we will drop the index $\omega$ in the
following and proof by contradiction.

We will use induction on $k$. Suppose that this is not true for the
first packet, i.e., $s_1>s_1'$. In $\mathcal{S}'$, due to maximal
scheduling, each time slot from $t_1'+1$ to $s_1'-1$ is occupied by
the neighbors of $i$ (which is why the packet in $i$ has not been
served yet). Note that we always have\vspace{-.05in}
\begin{equation}
\label{eqn: Ds>Ds'}D_i(n)+\sum_{j\in N_i}D_j(n)\geq
D_i'(n)+\sum_{j\in N_i}D_j'(n),
\end{equation}\vspace{-.15in}

\noindent for any $n\geq 1$ since both systems are work-conserving
and the links in $\mathcal{S}'$ have edges between them (they form a
clique), even when $\mathcal{S}$ does not. Thus from $s_1>s_1'$, we
have $D_i(s'_1-1)=D_i'(s'_1-1)=0$ and\vspace{-.07in}
\begin{eqnarray*}
\sum_{j\in N_i}D_j(s'_1-1)\geq \sum_{j\in N_i}D_j'(s'_1-1),
\end{eqnarray*}\vspace{-.15in}

\noindent following Eqn. (\ref{eqn: Ds>Ds'}). Therefore, we have
\vspace{-.07in}
\begin{eqnarray}
0\stackrel{(a)}{=}\sum_{j\in N_i}Q_j'(s_1'-1)&=&\sum_{j\in N_i}A_j'(s_1'-1)-\sum_{j\in N_i}D_j'(s_1'-1)\nonumber\\
&\geq&\sum_{j\in N_i}A_j(s_1'-1)-\sum_{j\in N_i} D_j(s_1'-1)\nonumber\\
&=&\sum_{j\in N_i}Q_j(s_1'-1)\stackrel{(b)}{>}0.\label{eqn:
delay_proof}
\end{eqnarray}\vspace{-.15in}

\noindent Equality $(a)$ is explained as follows. Since $i$
transmits in $\mathcal{S}'$ at time $s'_1$, no other neighbor is
contending for that slot (neighbors have higher priority). It must
be that all the neighbors have empty queues in the previous slot.
The inequality $(b)$ is explained as follows. Since $i$ does not
transmit in time slot $s_1'$ in $\mathcal{S}$ (we know it transmits
at $s_1>s_1'$), that must be due to some transmitting neighbor
occupying that slot. Since arrivals occur at the end of each slot,
that neighbor must have a nonempty queue in the previous slot
$s_1'-1$. Since this is a contradiction, we have proved the case for
$k=1$.

Now suppose it is true for up to $k-1$ packets. For the $k$th
packet, any time slot between $t'_k+1$ to $s_k'-1$ is occupied by
the neighbors in $\mathcal{S}'$. If $s_k>s_k'$, we have $s_{k-1}\leq
s_{k-1}'< s_k'<s_k$, and
\begin{equation}
D_i(s_k'-1)=D_i'(s_k'-1)=k-1
\end{equation}
Similar to (\ref{eqn: delay_proof}), we get\vspace{-.07in}
\begin{eqnarray*}
0=\sum_{j\in N_i}Q_j'(s_k'-1)&=&\sum_{j\in N_i}A_j'(s_k'-1)-D_j'(s_k'-1)\\
&\geq&\sum_{j\in N_i}A_j(s_k'-1)-D_j(s_k'-1)\\
&=&\sum_{j\in N_i}Q_j(s_k'-1)>0,
\end{eqnarray*}\vspace{-.15in}

\noindent i.e., a contradiction. Thus the lemma
holds.\hfill$\blacksquare$

Having shown the the dominance of $\mathcal{S}'$, we next analyze
the delay performance of link $i$ in $\mathcal{S}'$ in the following
lemma.

\emph{Lemma 4:} For link $i$ in system $\mathcal{S}'$, we
have\vspace{-.07in}
\begin{equation}
\lim_{x\rightarrow\infty}{-1\over x}\log\Pr(Q'_i(t)\geq
x)=\theta_i^*
\end{equation}\vspace{-.15in}

\noindent where $\theta_i^*$ is the largest root of the following
equation\vspace{-.07in}
\begin{equation}\label{eqn: delay_theta}
\Lambda_i(\theta)+\inf_{0\leq u\leq\theta}[\sum_{j\in
N_i}\Lambda_j(u)-u]=0
\end{equation}
\vspace{-.2in}

\emph{Proof:} We construct a system $\mathcal{S}''$ equivalent to
$\mathcal{S}'$, consisting of two nodes, where node $1$ corresponds
to link $i$ and node $2$ corresponds to the links in $N_i$. Thus,
the queue of node 2 is the sum of the queues of link $i$'s
neighbors. We assume that the arrival process to node 2 is equal to
the sum arrivals in $N_i$, and that there is an edge between 1 and
2. During the scheduling, we assign higher priority to node $2$.
Note that from link $i$'s perspective, $\mathcal{S}'$ and
$\mathcal{S}''$ yield the same queueing dynamics.

For a 2-node system, according to \cite{bertsimas99}, the delay
exponent $\theta^*$ for queue 1 is given by the largest root of the
equation
\begin{equation}
\Lambda_1(\theta)+\inf_{0\leq u\leq\theta}[\Lambda_2(u)-u]=0.
\end{equation}
Thus the lemma holds from the fact that the arrival process of node
$2$ is $A_2(n)=\sum_{j\in N_i} A_j(n)$.\hfill$\blacksquare$

From the above discussions, we have the following theorem:

\emph{Theorem 5:} In a network where the arrival processes satisfy
{\bf A1-A3}, we have $\Theta_{\min}\subseteq\Theta_\pi$ for any
maximal scheduler $\pi$, where
$\Theta_{\min}=\{\theta\in\mathbb{R}_n^+:\theta\prec\theta^*\}$ and
$\theta^*_i$ is the largest root of Eqn. (\ref{eqn: delay_theta}).

This theorem provides a lower bound on the delay region of any
maximal scheduler.

\subsection{Delay-Aware Maximal Scheduling}

Having obtained the asymptotic limit of the worst case maximal
scheduler, we next consider improving the guarantees through proper
priority assignment. In the following we assume that traffic
parameter $\Lambda_i(\theta)$ for each link $i$ is available.
$\Lambda_i(0)=a_i$, so that $\Lambda_i(\theta)$ is more detailed
than $a_i$. First we generalize Theorem 5 to prioritized maximal
schedulers.

\emph{Lemma 5:} The delay region
$\Theta_p=\{\theta\in\mathbb{R}_+^n: \theta\prec\theta^*\}$ can be
achieved by the prioritized maximal scheduler $\pi_p$, where
$\theta_i^*$ is the largest root of the following
equation\vspace{-.05in}
\begin{equation}
\Lambda_i(\theta)+\inf_{0\leq u\leq\theta}[\sum_{j\in
N_i}\Lambda_j(u){\bf 1}(p_i-p_j)-u]=0
\end{equation}\vspace{-.1in}

\emph{Proof:} Note that the neighbors in $N_i$ with lower priorities
than link $i$ are invisible to link $i$. Therefore, we create a
dominating system $\mathcal{S}'$ consisting of a clique formed by
link $i$ and its neighbors that have higher priorities than link
$i$. Using a similar argument, one can show that Lemma 3 holds.
Therefore, Lemma 5 holds, following Lemma 4.\hfill$\blacksquare$

Similar to $\mathcal{A}$, we denote $\Theta=\cup_p\Theta_p$, i.e.,
the delay region guaranteed by \emph{any} constant priority
scheduler. We are interested in computing a proper priority $\bf p$,
when given a QoS constraint in the form of the vector $\theta$. We
will show this can be solved by an algorithm adapted from
Stable-Priority, under the assumption that $\theta\in\Theta$, by
showing that $\Theta$ has a structure similar as $\mathcal{A}$.

\emph{Lemma 6:} $\Theta=\Theta'$, where $\Theta'$ is any delay
region in $\mathbb{R}^n_+$ that satisfies the following conditions:
\begin{enumerate}
\item $\Theta'$ is coordinate-convex;
\item For any nonzero $\theta\in\Theta'$, there exists $1\leq i\leq
n$ such that $\theta_i>0$ and $\theta_i$ is less than the largest
root of the following equation\vspace{-.05in}
\begin{equation}\label{eqn: theta_i*}
\Lambda_i(\theta)+\inf_{0\leq u\leq\theta}[\sum_{j\in
N_i}\Lambda_j(u){\bf 1}(\theta_j)-u]=0
\end{equation}\vspace{-.15in}
\end{enumerate}
where ${\bf 1}(\cdot)$ is the indicator function in (\ref{eqn:
indicator}).

\emph{Proof:} Suppose $\theta\in\Theta$. Then, $\theta\in\Theta_{\bf
p}$ for some $\bf p$. Therefore property (1) holds following the
fact that $\Theta_{\bf p}$ is coordinate-convex. We can assume that
$\bf p$ is such that the links with zero delay exponents have the
lowest priorities. Next, consider the lowest priority link $i$ with
$\theta_i>0$ according to $\bf p$. We have $\theta_i<\theta_i^*$
where $\theta_i^*$ is the largest root of the following equation
\begin{equation}
\Lambda_i(\theta_i)+\inf_{0\leq u\leq\theta_i}[\sum_{j\in
N_i}\Lambda_j(u){\bf 1}(p_i-p_j)-u]=0.
\end{equation}
Thus property (2) holds from the fact that ${\bf 1}(p_i-p_j)={\bf
1}(\theta_j)$ for all $j\in N_i$. Hence we have
$\Theta\subseteq\Theta'$.

Now suppose $\theta^{(1)}\in\Theta'$. We first assign the lowest
priorities to any link $i$ such that $\theta^{(1)}_i=0$. According
to property (2), we can find link $i_1$ such that
$0<\theta^{(1)}_{i_1}<\theta^{(1)*}_{i_1}$ where
$\theta^{(1)*}_{i_1}$ is described in Eqn. (\ref{eqn: theta_i*}). We
assign ${i_1}$ the current lowest priority available and set
$\theta^{(1)}_{i_1}=0$ to get a new vector $\theta^{(2)}\in \Theta$.
Repeat the above process until all the links are assigned
priorities. Denoting the resulting priority vector as $\bf p$, it is
easy to check that $\theta^{(1)}\in\Theta_{\bf p}$, and hence
$\Theta'\subseteq\Theta$. Therefore the lemma
holds.\hfill$\blacksquare$

Based on the above lemma, we can construct a priority assignment
algorithm, such that whenever $\theta\in\Theta$, the algorithm will
output a satisfying priority. This is the delay-equivalent of
Stable-Priority algorithm, which was constructed using Lemma 2.
\vspace{-.1in}
\begin{algorithm}[H]
\begin{algorithmic}[1]
\caption{Delay-Priority $(G_c,\theta)$}

\FOR{$k=1$ to $n$}

\FOR{$i\in V_c$}

\STATE Compute the largest root $\theta_i^*$ of the equation:\\
$\Lambda_i(\theta)+\inf_{0\leq u\leq\theta}[\sum_{j\in
N_i}\Lambda_j(u){\bf 1}(\theta_j)-u]=0$

\ENDFOR

\STATE $s\leftarrow\min\{i\in V_c:\theta_i<\theta_i^*\}$

\STATE $p_s\leftarrow n+1-k$

\STATE $\theta_s\leftarrow0$

\ENDFOR

\RETURN $\bf p$
\end{algorithmic}
\end{algorithm}
\vspace{-.15in}

We conclude with the following theorem, which is the delay-analogue
of Theorem 3.

\emph{Theorem 6:} If $\theta\in\Theta$, Delay-Priority will generate
a priority $p$ such that the $\theta\in\Theta_p$.

\section{Conclusion}
\label{sec_conclusion}

This paper considered the prioritized maximal scheduling problem in
multi-hop networks for arbitrary correlated arrival processes. We
first considered the random priority case and showed that one can
achieve the optimal stability region with i.i.d priorities. Then we
focused on the constant priority case and proposed a priority
assignment algorithm, which, combined with maximal scheduling, can
achieve an efficiency ratio, which is the same as that achieved by
the state of art LQF scheduling. We also analyzed the delay
performance of maximal schedulers in the large deviations regime,
assuming independent arrival processes, and proposed a delay-aware
prioritized maximal scheduling algorithm.

%

\appendix
\emph{Proof of Lemma 1:} Due to space limit, we only give a outline
of the proof, for a detailed discussion about fluid limits please
refer to \cite{chaporkar08} and \cite{dai00}.

\emph{1) The existence of fluid limits:} Note that the support of
functions $A_i(n), D_i(n)$ and $Q_i(n)$ is $\mathbb{N}$, we extend
it to $\mathbb{R}_+$ using linear interpolation, which results in
continuous functions (e.g., $A_i(t), t\in \mathbb{R}_+$). For any
sample path $\omega$, define a family of functions as\vspace{-.1in}
\begin{equation}\label{eqn: scaling}
f^r(t,\omega)={f(rt,\omega)\over r}
\end{equation}\vspace{-.2in}

\noindent where $f(\cdot)$ could be $A_i(\cdot), D_i(\cdot)$ or
$Q_i(\cdot)$. Since both the arrivals and the departures in each
time slot are bounded, and by linearity in (\ref{eqn: queueing}),
the functions defined in (\ref{eqn: scaling}) are Lipschitz
continuous with the same Lipschitz constant (irrespective of $r$),
and hence equi-continuous on any compact interval. Since they are
also uniformly bounded on $[0, t]$, according to Arzela-Ascoli
theorem, in any $[0,t]$, there exists a subsequence $r_{n_k}$ and
continuous functions $\bar{A}_i(\cdot), \bar{D}_i(\cdot)$ and
$\bar{Q}_i(\cdot)$, such that (uniform convergence)\vspace{-.05in}
\begin{eqnarray*}
\lim_{k\rightarrow\infty}\sup_{\tau\in[0,t]}|f_i^{r_{n_k}}(\tau,\omega)-\bar{f}_i(
\tau,\omega)|=0
\end{eqnarray*}\vspace{-.12in}

\noindent where $f(\cdot)$ could be $A(\cdot), D(\cdot)$ or
$Q(\cdot)$. Any $(\bar{A}_i(\cdot),
\bar{D}_i(\cdot),\bar{Q}_i(\cdot))$ satisfying the above is defined
as a fluid limit.

\emph{2) Properties of fluid limits:} For any fluid limit, we have
$\bar{A}_i(t)=a_it$ w.p.1, since $A_i(n)$ satisfies SLLN,
and\vspace{-.07in}
\begin{equation}
\bar{Q}_i(t)=\bar{Q}_i(0)+a_it-\bar{D}_i(t)
\end{equation}\vspace{-.2in}

\noindent for all $i\in V_c$. The network is stable if, for any
fluid limit with $\bar{Q}_i(0)=0, \forall i$, we have
$\bar{Q}_i(t)=0$ for all $i$ and all $t>0$.

\emph{3) Stability proof of Lemma 1:} Define the Lyapunov
function\vspace{-.05in}
\begin{equation}
L_i(t)=Q_i(t)+\sum_{j\in N_i/S_i}Q_j(t)
\end{equation}\vspace{-.11in}

\noindent To prove the stability, it is sufficient to prove that if
$\bar{L}_i(0)=0$ for all $i$, we have $\bar{L}_i(t)=0$ for all $i$
and $t>0$. Suppose that this is not true, then there exists $i\in
V_c$ and $t>0$ such that\vspace{-.07in}
\begin{eqnarray}
\bar{L}_i(t)&=&\max_{\tau\in[0,t]}\bar{L}_i(\tau)\label{eqn: lyapunov}\\
\bar{Q}_i(t)&=&x>0,
\end{eqnarray}\vspace{-.15in}

\noindent for details see the proof for (25) and (26) in
\cite{chaporkar08}. Since $\bar{Q}_i(t)$ is a uniformly continuous
function, there is a $t'<t$ such that $\bar{Q}_i(\tau)\geq x/2$ for
$\tau\in[t',t]$. Note that since $\bar{Q}_i(t)$ is a fluid limit,
there is a subsequence $r_{n_k}$ such that\vspace{-.07in}
\begin{equation}
\bar{Q}_i(\tau)=\lim_{k\rightarrow\infty}{Q_i(r_{n_k}\tau)\over
r_{n_k}}\geq{x\over 2}
\end{equation}\vspace{-.15in}

\noindent for every $\tau\in[t',t]$. Particularly, for large enough
$k$, we have\vspace{-.07in}
\begin{equation}
Q_i(r_{n_k}\tau)\geq{x\over 4}r_{n_k}\geq 1\qquad\forall
\tau\in[t',t]
\end{equation}\vspace{-.15in}

\noindent i.e., the queue of link $i$ is nonempty during
$[r_{n_k}t', r_{n_k}t]$. Therefore we have\vspace{-.07in}
\begin{eqnarray*}
L_i(r_{n_k}t)-L_i(r_{n_k}t')=\sum_{j\in
\{i\}\cup(N_i/S_i)}[A_j(r_{n_k}t)-A_j(r_{n_k}t')]\\
-\sum_{j\in \{i\}\cup(N_i/S_i)}[D_j(r_{n_k}t)-D_j(r_{n_k}t')]
\end{eqnarray*}\vspace{-.15in}

\noindent Due to the randomized priority scheduling, since queue $i$
is non-empty, if $i$ has higher priority than any link in $S_i$ in a
certain slot, we can guarantee one departure in $\{i\}\cup
(N_i/S_i)$. Since the priorities are i.i.d across time slots, by
SLLN\vspace{-.07in}
\begin{eqnarray*}
&&\lim_{k\rightarrow\infty}{\sum_{j\in
\{i\}\cup(N_i/S_i)}[D_j(r_{n_k}t)-D_j(r_{n_k}t')]\over
r_{n_k}}\\&\geq&(t-t')\Pr(\textrm{queue $i$ has local highest
priority})\\&=&\Pr(p_i<p_j,\forall j\in S_i)(t-t').
\end{eqnarray*}\vspace{-.2in}

\noindent Note that if,\vspace{-.1in}
\begin{eqnarray*}
&&\lim_{k\rightarrow\infty}{\sum_{j\in
\{i\}\cup(N_i/S_i)}[A_j(r_{n_k}t)-A_j(r_{n_k}t')]\over r_{n_k}}\\
&=&(a_i+\sum_{j\in N_i/S_i}a_j)(t-t')\qquad\textrm{(SLLN)}\\
&<&\Pr(p_i<p_j,\forall j\in S_i)(t-t')
\end{eqnarray*}\vspace{-.22in}

\noindent then we have\vspace{-.07in}
\begin{eqnarray*}
\bar{L}_i(t)-\bar{L}_i(t')=\lim_{k\rightarrow\infty}
L_i^{r_{n_k}}(t)-L_i^{r_{n_k}}(t')<0
\end{eqnarray*}\vspace{-.2in}

\noindent which contradicts Eqn. (\ref{eqn: lyapunov}). Therefore
the stability of the network is proved and the lemma
holds.\hfill$\blacksquare$

\end{document}